\documentstyle[11pt,paspconf]{article}

\begin{document}

\title{Abundance Profiles in Disk Galaxies from Nebulae\footnotemark[1]}
\footnotetext[1]{Invited review presented at the workshop on ``Abundance Profiles: Diagnostic Tools For Galaxy History'', held October 12-15, 1997, at Universit{\'e} Laval, Qu{\'e}bec City, Qu{\'e}bec.}
\author{R.B.C. Henry} 
\affil{Department of Physics \& Astronomy, University of Oklahoma, Norman, OK  73019 USA}

\begin{abstract}

Galactic and extragalactic abundance studies of H~II regions, planetary
nebulae, and supernova remnants are compiled and their implications briefly
reviewed.  Galaxy global metallicity appears to be related directly to
total mass and morphology, while gradient slopes are uncorrelated
with these parameters, although  barred spirals tend to have flatter
profiles than non-barred spirals.  N/O behavior is consistent with
primary, then secondary, production of N as metallicity increases, and
empirical evidence for a metallicity-sensitive C yield is seen.
Metallicity-invariant primary production mechanisms are suggested for
S, Ne, Ar, and O.

\end{abstract}


\section{Introduction}

Abundance measurements of C, N, O, Ne, S, and Ar in galactic and
extragalactic H~II regions, planetary nebulae, and supernova remnants
represent a major source of information about elemental levels in the
interstellar media of spiral disks.  Measured from the ground in most
cases, the strengths of the numerous emission features produced by
these objects can be converted in a straightforward way to ionic and
elemental abundances.  When the abundances for nebular objects within a
single galaxy are compiled, several correlations are shown to exist
either between interstellar metallicity and galactocentric distance,
i.e.  an abundance gradient, or between pairs of abundance ratios.
Since the former is directly linked to matter distribution and star
formation rates, and the latter depends on the IMF and stellar yields,
complete abundance profile information for a galactic disk provides
important constraints on galactic chemical evolution models and hence
our global understanding of how elemental composition changes within a
galaxy as a function of time and location.  The purpose of this review
is to provide a summary of extant data pertaining to nebular abundance
profiles in disk galaxies along with brief interpretations.  Readers are referred to other papers in this volume for more detailed theoretical explorations of abundance gradients.

\section{Metallicity Profiles of Spiral Disks}

Because oxygen is readily accessible spectroscopically, its
abundance provides a convenient tracer of metallicity
distribution in a galactic disk.  I begin by discussing the Milky
Way oxygen profile and then follow up with a summary of general
results for a large number of other spirals.

Data for oxygen in the Milky Way disk were taken from the
following papers:  Shaver et al. (1983), the first major survey of
abundances in galactic H~II regions; V{\'i}lchez \& Esteban
(1996), a focused study of H~II regions at large galactocentric
distances; and Maciel \& K{\"o}ppen (1994), where a large number
of galactic type~2 planetary nebulae were used to map abundances
in the disk.  Abundances were generally derived in these papers by combining
measured line strengths directly with atomic data and ionization
correction factors as described in Osterbrock (1989).  

Figure 1 shows 12+log(O/H) versus galactocentric distance in kpc
for the Milky Way, where symbol shape indicates the data source. 
\begin{figure} 
\plotfiddle{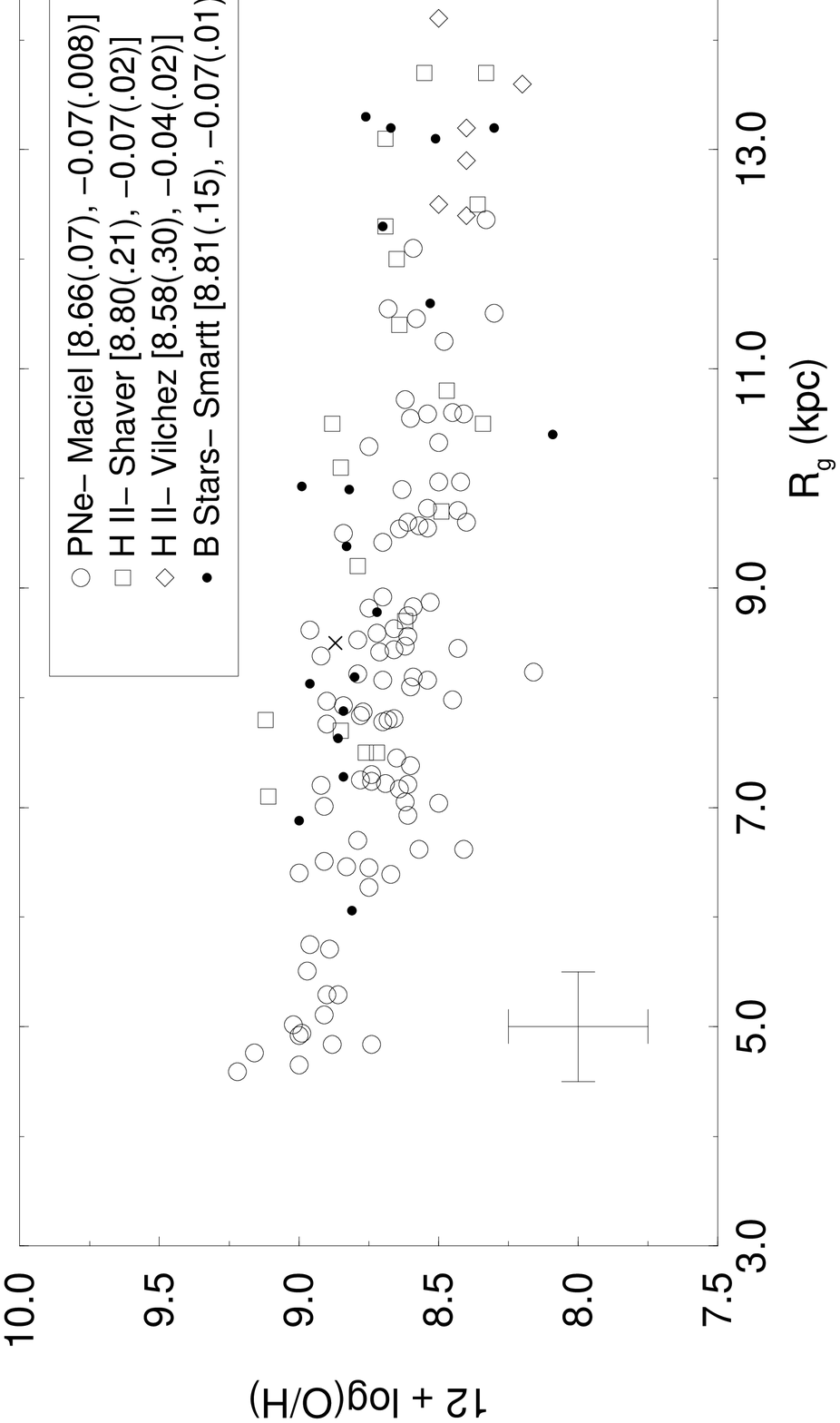}{50truemm}{270}{35}{35}{-200}{180}
\caption{12+log(O/H) versus galactocentric distance in kpc for the four
data sets indicated in the legend, where parameters (and
uncertainties) for least squares fits
for each data set are included (see text).} 
\end{figure}
Also included are the recent B star results from Smartt \&
Rolleston (1997).  Numbers in square brackets in the legend give
12+log(O/H) at the sun's distance (8.5~kpc) and the slope in
dex/kpc, respectively, for least squares fits.  Parameter uncertainties are given in parentheses.  The sun's
position (Grevesse \& Noels 1993) is indicated with an `x'.  Error
bars here and in subsequent figures show typical uncertainties.

Besides the presence of a negative
metallicity gradient, Fig.~1 implies the following.  (1)~The
galactic oxygen gradient flattens beyond 10kpc, according to the
V{\'i}lchez H~II region data.  (2)~The B star oxygen
profile is consistent with H~II region results inside of 10kpc
and shows no flattening beyond 10kpc.  (3)~The oxygen abundances
in planetary nebulae are systematically less than in H~II regions
at corresponding distances by roughly 0.14~dex, qualitatively
consistent with the idea that planetary nebulae represent an
older, less metal-rich population than H~II regions.

Turning now to the consideration of spiral galaxies generally,
large surveys of O/H in extragalactic H~II regions include those
of McCall (1982; 40 galaxies), Vila-Costas \& Edmunds (1992; 32
galaxies), and Zaritsky, Kennicutt, \& Huchra (1994; 39 galaxies).
 Detailed results for O/H in individual spirals can be found in
these papers.  To show general findings here I have extracted
characteristic abundances\footnote[2]{The characteristic abundance 
is the abundance at 0.4r$_o$ as
determined by a least squares fit to the data, where r$_o$ is the
isophotal radius at $\sigma_B=25$ mag arcsec$^{-2}$.} 
and gradient slopes from Zaritsky et al. and present them in Figure~2 
\begin{figure}
\plotfiddle{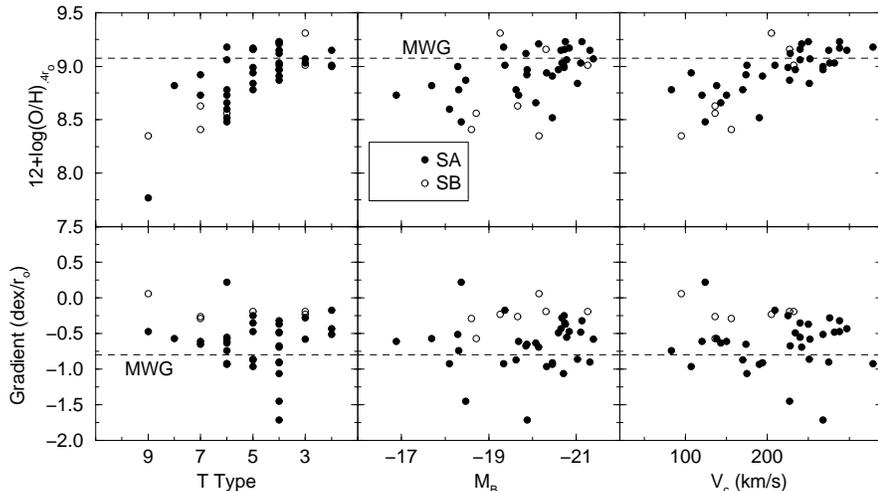}{55truemm}{270}{50}{50}{-175}{260}
\caption{Characteristic abundances and gradient slopes from
Zaritsky et al. (1994) plotted against morphological T Type,
absolute blue magnitude, and circular velocity.  Regular and
barred spirals are shown with filled and open circles,
respectively.  Parameters for the Milky Way shown in Fig.~1 for
Shaver's data are indicated with dashed lines.}
\end{figure}
as functions of galaxy morphological type (T type), absolute blue
magnitude M$_B$, and maximum circular velocity V$_c$ in km/s.  All three of
these independent parameters are indicators of galaxy mass, where
smaller T type indices, more luminous integrated blue magnitudes, and larger
rotational velocities generally correspond with more massive spirals.
Gradient slopes are shown in
dex/r$_o$.  Normal and barred spirals are shown separately using
filled and open symbols, respectively.
Parameters from the Shaver data for the Milky Way are indicated in Fig.~2 
with dashed lines, where I have adopted 11.5kpc for the isophotal radius 
(de~Vaucouleurs \& Pence 1978).  I note that while isophotal radius is 
employed as the normalization standard here, use of effective
radius (the half light radius) or kiloparsecs produces similar
results.  The choice of normalization standard is
discussed by Garnett in this volume.

Two points are implied by Fig.~2:
(1)~Characteristic abundances increase with galaxy mass, while
gradient slopes are uncorrelated with this parameter; and
(2)~Characteristic
abundances in normal and barred spirals are indistinguishable,
but
barred spirals appear to have flatter gradients.  Both of these results
have been noted previously.  Garnett \& Shields (1987) plotted
characteristic O/H values against galaxy mass for numerous spirals and
found a direct correlation between these two
parameters, while Pagel et al. (1979) first suggested that
barred spirals may have flatter gradients.

\section{Abundance Ratios of Heavy Elements}

While metallicity patterns are apparently related closely to global
galaxy properties, heavy element ratios such as N/O are expected to
reflect characteristics of the IMF, stellar yield patterns, 
and star formation history.  Papers
on heavy element abundance ratios for the Milky Way included in this
review are: Shaver et al. (1983; N/O, S/O, Ne/O, and Ar/O), V{\'i}lchez
\& Esteban (1996; N/O, S/O), Simpson et al. (1995; N/O, S/O, Ne/O),
Maciel \& K{\"o}ppen (1994; Ne/O, S/O, Ar/O), and Fesen, Blair, \&
Kirshner (1985; N/O, S/O).  Ratios for extragalactic H~II regions are
taken from papers by Thurston, Edmunds, \& Henry (1996; N/O),
Kobulnicky \& Skillman (1996; N/O), Garnett (1989; S/O), and Garnett et
al. (1995, 1997; C/O).  Note that Simpson et al. used lines in the IR
to infer their abundances.  All papers considered H~II regions with
the exception of those by Fesen et al.  and Maciel \& K{\"o}ppen
which studied
supernova remnants and planetary nebulae, respectively.

Figure 3 shows log(N/O)
versus 12+log(O/H) for both the Milky Way and extragalactic
objects.  Milky Way data are shown with
symbols whose shapes are interpreted in the legend of Fig.~4.
\begin{figure}
\plotfiddle{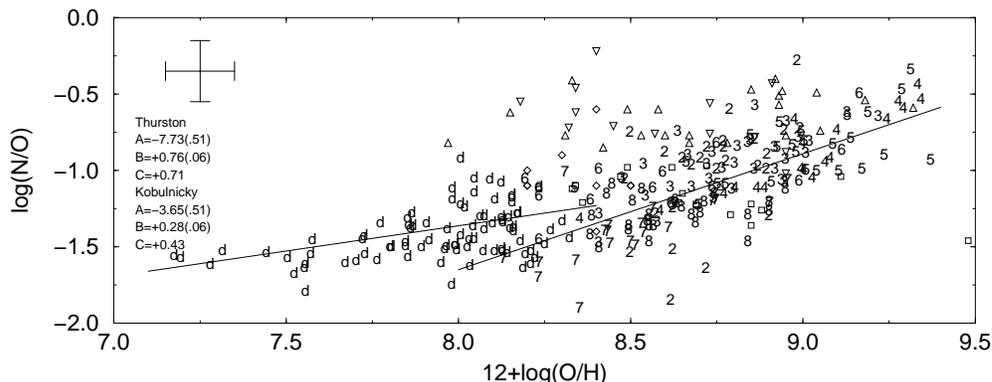}{40truemm}{270}{40}{40}{-200}{215}
\caption{log(N/O) versus 12+log(O/H) for Milky Way data (symbol
shapes interpreted in Fig.~4) along with the extragalactic data
of Thurston et al. (1996; integers indicating T type) and Kobulnicky \& Skillman (1996;
`d').  Parameters (and uncertainties) 
for least squares fits are also shown for these two extragalactic data sets.}
\end{figure}
Notice the segregation in the Milky Way data 
of the Simpson and Fesen points from the Shaver and Maciel points, perhaps 
due to the differences in the way these ratios were measured.
T~Type values are used as symbols for the Thurston et al. spirals, 
while `d' designates the Kobulnicky \& Skillman 
dwarf galaxies.  Parameters for least squares fits are indicated in the figure 
for the two extragalactic data sets only, where A=y-intercept, B=slope, and 
C=correlation coefficient.  
Because of the scatter in the Milky Way data, 
the following analysis applies only to the extragalactic N/O studies.  

The flatter N/O behavior seen at 12+log(O/H)$<$8.0 can be ascribed to
the dominance of
primary nitrogen production where CNO cycling is fed with carbon and
oxygen derived directly from triple alpha processing within the same
star.  Thus, N/O is relatively independent of the star's metallicity.  The
steeper slope in N/O at higher metallcities is probably related to
metallicity-sensitive secondary nitrogen production in which significant 
amounts of carbon and
oxygen already present in the star enters into the
CNO cycle.  [See Vila-Costas \& Edmunds
(1993) and Thurston et al. (1996) for further discussion.]  
Finally, note the tendency for early type spirals (low integer 
values) to have systematically higher log(N/O) for the same 12+log(O/H).

C/H, S/H, Ne/H, and Ar/H are plotted logarithmically
against 12+log(O/H) in Fig.~4, where the legend
connects symbol shape with data source, and the parameters (and
uncertainties) defined as in Fig.~3 for the least squares linear fits are
shown in each panel.  Note that open and filled symbols designate
Milky Way and extragalactic H~II regions, respectively.

\begin{figure}
\plotfiddle{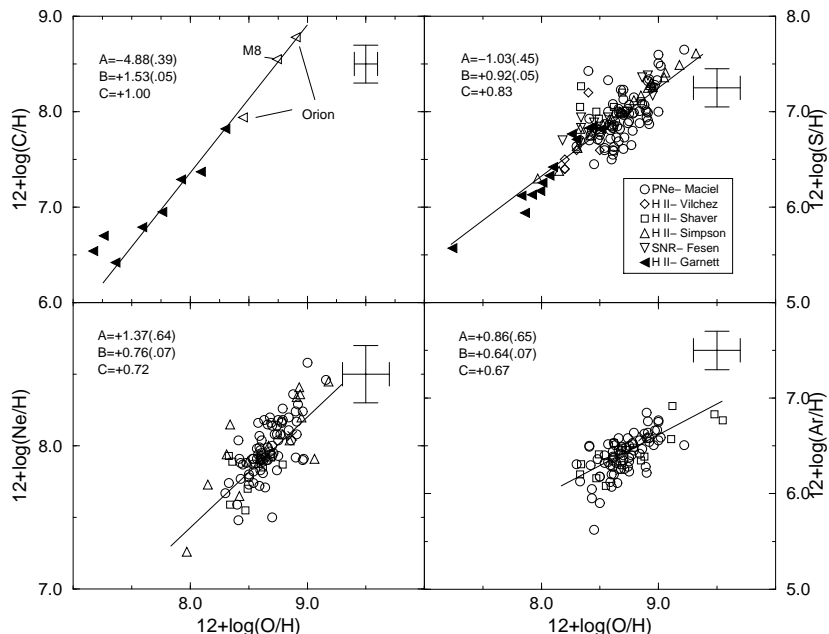}{75truemm}{270}{50}{50}{-175}{275}
\caption{Logarithmic C/H, Ne/H, S/H, and Ar/H abundance ratios
plotted against O/H for the samples indicated in the legend.  Fit
parameters (and uncertainties) are shown for each case.}
\end{figure}

The top left panel of Figure~4 shows C/H values from Garnett et al. (1995; 1997) for
extragalactic H~II regions, along with results for galactic
objects M~8 (Peimbert et al. 1993) and the Orion Nebula (Walter
et al. 1992).  
The two results for Orion stem from the
assumption of line-of-sight temperature fluctuations in the case
of the higher value.  The two points at low O/H are for I~Zw~18.  
For purposes of the fit, the Orion collisional line abundance
and the two points for I~Zw~18 were omitted.  A clear correlation exists 
between C and O, and with a slope significantly greater than unity, a 
metallicity-sensitive production mechanism for carbon is suggested.

Another strong correlation, assisted considerably by the extragalactic
data, exists between S/H and O/H (top right panel).  It is interesting
to note the greater scatter for planetary nebula data than for H~II
region data.  In all but the Simpson et al. work, S abundances were
determined by combining S$^+$ abundances with an ionization correction
factor.  However, the excitation level of planetary nebulae is
generally higher than in H~II regions, so planetaries should have
proportionately less S$^+$, and this method will be less accurate for
them.  The strong correlation shown in this plot, along with a slope
close to unity, implies that S/O is generally constant.  The average
log(S/O) value is -1.76, close to the solar value of {-1.66} (Grevesse
\& Noels 1993).  This result agrees with Garnett (1989), although
D{\'i}az (1989) found a negative correlation between log(S/O) and
12+log(O/H).  In addition to ionization correction factor uncertainties, sulfur abundance studies also suffer from the effects of severe telluric absorption on the IR [S~III]
9069{\AA} and 9532{\AA} lines.

Ne/H and Ar/H appear to be correlated with O/H (bottom two
panels),  although the slopes are less than unity.  Nevertheless,
average ratios of $log(Ne/O)=-0.70$ and $log(Ar/O)=-2.30$, compare
favorably with solar values of {-0.80} and {-2.27}
(Grevesse \& Noels 1993), respectively.

Tight correlations such as those between S, Ne, Ar, and O suggest
that these elements are produced by primary nucleosynthesis in
massive stars with yields that are invariant with metallicity. 
This issue was explored in the case of Ne and O in particular by
Henry (1989), who found the interstellar Ne/O ratio to be
constant in several different galaxies.

\section{Assorted Issues}

{\it The effects of bars on abundance gradients} has recently
been studied extensively in particular by Martin, Roy, Walsh,
Belley, and Julien.  The bottom three panels of Fig.~2 
show the tendency for barred spirals to possess shallower
gradients.  The paper by Martin \& Roy (1994) further relates
gradient slope to bar strength, a quantity which measures bar
ellipticity.  They find direct relations between the slope of the
oxygen abundance gradient of a barred spiral and the galaxy's
bar strength (ellipticity) and length.  This empirical
result is consistent with radial flow models of chemical
evolution in which the presence of a bar enhances large-scale
mixing over the galaxy's disk, damping radial abundance
variations.

{\it A negative vertical gradient in O/H in the Milky Way} is
suggested by planetary nebula studies.  Abundance data compiled
by Kaler (1980) for PNe ranging in height above the disk from
less than 0.4~kpc to greater than 1~kpc show a decrease in O/H
with increasing height above the plane.  A comparison of more
recent studies of PNe close to the plane (Perinotto 1991),
greater than 300pc above the plane (Cuisinier et al. 1996), and
in the halo (Howard \& Henry 1997) shows averages of 12+log(O/H)
for these three samples of 8.68, 8.52, and 8.02 respectively,
qualitatively consistent with Kaler.

Thorough tests for {\it azimuthal gradients in spiral disks} have
yet to be carried out.  One example of apparent O/H asymmetry is
discussed by Kennicutt \& Garnett (1996) in their study of M101. 
They find that H~II regions located along a spiral arm southeast
of the major axis have a lower oxygen abundance by 0.2-0.4~dex
compared with H~II regions on the opposite side.

{\it Global metallicities in low surface brightness galaxies} are
generally found to be subsolar by roughly a factor of three, according
to McGaugh (1994), indicating that these galaxies evolve very slowly
and form few stars during a Hubble time.  Apparently, they also
lack detectable gradients.
This, despite the fact that
these objects are similar in mass and size to prominent spirals
defining the Hubble sequence.
McGaugh suggests that a galaxy's environment and surface
mass density are more relevant to galaxy evolution than gross size.

{\it Effects of cluster environment} on the chemical evolution of
galaxies has been investigated by Skillman et al. (1996), who studied
oxygen profiles in several Virgo spirals representing a range in
H~I deficiency (taken as a gauge of cluster environmental interactions).
Their results imply
that global metal abundances in disks tend to be higher in stripped
galaxies, presumably because reduced infall of metal-poor H~I gas means
less dilution of disk material.  Henry et al. (1996 and
references therein) investigated metallicity
and heavy element abundance ratios (N/O, S/O) in three cluster spiral disks
with normal H~I and found no clear signatures of environmental
effects.  Thus, cluster environment alone is apparently not a
sufficient condition for altered chemical evolution.

{\it The mathematical form of abundance profiles in spiral
disks,} has been
investigated recently by Henry \& Howard (1995), who fit
line strength behavior over the disks of M33, M81, and M101
using photoionization models.
Their best fits for O/H versus galactocentric distance
were produced using exponential profiles, although
power law forms could not be ruled out.  However, linear
profiles poorly reproduced the observations.  Henry and Howard
also concluded that there is currently no strong observational
case for gradient flattening in the outer parts of some disks,
although such flattening has been proposed by several authors
(see Moll{\`a} et al. 1996). 

\section{Summary}

Since the conference review on
this same subject by D{\'i}az (1989) both the average number of
observed and analyzed H~II regions per galaxy and the number of
galaxies sampled have steadily risen.  Also, a larger number
of elements is now being studied.  The following points seem solid:
(1)~Global metallicity in spirals is influenced by galaxy mass. Evidence in
Vila-Costas \& Edmunds (1992) also suggests that interstellar
metallicity scales directly with total surface density, perhaps
through a star formation rate which is a function of local
density, and thus an abundance gradient merely traces matter
distribution.  (2)~N/O behavior with metallicity clearly suggests
a primary then secondary origin of N with chemical evolution.
(3)~Finally, constant ratios of S/O, Ne/O, and Ar/O are
ubiquitous and appear to suggest primary nucleosynthesis in
massive stars with metallicity-insensitive yields.

While significant progress has been made over the past decade, more
questions remain to be addressed.  Is the abundance scatter at a
specific galactocentric distance real or observational? Why do
collisional and recombination lines give abundances which often differ
by as much as a factor of two?  Do gradients really flatten in outer
regions of disks?  Finally, there is the problem of measuring good
abundances in low excitation H~II regions, where
auroral line electron temperature determinations are difficult/impossible
to obtain, preventing adequate probes of early spirals as well as the inner disks of late ones.  Clearly, there is still plenty of work to do.

\acknowledgments
I am grateful to Dennis Zaritsky for sharing his abundance data, Reggie Dufour for his stimulating comments, and the University of Oklahoma for assisting with travel support.

\end{document}